\documentclass[seceq]{ptptex}

\notypesetlogo                       

\title{
\hfill{\normalsize%
\vbox{\hbox{\rm DPNU-04-07}\hbox{\rm April, 2004}  }}\\
\vspace{0.2cm}
Vector Manifestation in Hot Matter~\footnote{
Talk given at the YITP workshop on ``Nuclear Matter 
under Extreme Conditions" (Matter03), Dec. 1-3, 2003, Kyoto, Japan.
This talk is based on the works done in 
Refs.~\citen{HY:VM,HY:PRep,HS:VMT,HS:VD,HKRS}.
}%
}

\subtitle{Formulation and Predictions} 

\author{
Masayasu \textsc{Harada}%
}

\inst{
Department of Physics, Nagoya University, Nagoya 464-8602, Japan
}

\abst{
In this write-up, I list the key ingredients for formulating
the vector manifestation in hot matter together with
several predictions made so far.
}

\begin{document}

\maketitle

\section{Introduction}
\vspace{-0.2cm}

The vector manifestation (VM) was proposed~\cite{HY:VM,HY:PRep}
as a novel manifestation in which 
the chiral symmetry is restored by 
the massless degenerate pseudoscalar meson 
and the vector meson as the chiral partner.
In Ref.~\citen{HS:VMT},
it was shown how the VM 
is formulated in hot matter using the effective field theory
for $\pi$ and $\rho$ based on the
hidden local symmetry (HLS)~\cite{BKUYY}.
In this write-up 
I shall 
list the key points for the formulation 
and several predictions.
For details please see the relevant references.

\section{Formulation of the vector manifestation in hot matter}
\vspace{-0.2cm}

The key ingredients to formulate the VM in hot matter are
the fixed point structure of the renormalization group equations
(RGEs) 
for the parameters in the HLS
and the {\it intrinsic temperature
dependences} of the parameters determined through the Wilsonian
matching~\cite{HY:WM}.
The coupled RGEs for the HLS gauge coupling $g$ and the
parameter $a$ have the fixed point 
characterized by $(g,a)=(0,1)$.
The intrinsic temperature dependence introduced through
the Wilsonian matching is
nothing but the signature that hadron has an internal structure 
constructed from the quarks and gluons.
This is similar to the situation where the coupling constants
among hadrons are replaced with the momentum-dependent form factor
in high energy region.
Thus the intrinsic thermal effects play more important
roles in higher temperature region, 
especially near the critical temperature.

The formulation of the VM in hot matter is roughly sketched as
follows~\cite{HS:VMT,HS:VD}:
The restored chiral symmetry implies that, 
at the critical temperature $T_c$,
the vector current correlator must agree with the axial vector 
current correlator.
The requirement of the equality between two correlators
implies that the bare $g$ and $a$ satisfy 
$(g_{\rm bare},a_{\rm bare})=(0,1)$.
Since $(g,a)=(0,1)$ is the fixed point of the RGEs for $g$ and $a$,
$(g_{\rm bare},a_{\rm bare})=(0,1)$ implies that 
$(g,a)=(0,1)$ is satisfied at any energy scale.
As a result, the quantum correction to the $\rho$ mass
as well as the hadronic thermal correction disappears at $T_c$
since they are proportional to the gauge coupling $g$.
The bare $\rho$ mass, which is also proportional to $g_{\rm bare}$,
vanishes at $T_c$.
These imply that the pole mass of the $\rho$ meson also vanishes
at $T_c$.

I would like to note 
that the VM in dense matter can be formulated
in a similar way~\cite{HKR}, 
where the {\it intrinsic density dependence}
plays an important role.

\section{Predictions of the vector manifestation in hot matter}
\vspace{-0.2cm}

There are several predictions of the VM in hot matter
made so far.

In Ref.~\citen{HKRS},
the vector and axial-vector susceptibilities were studied.
It was shown that the equality between two susceptibilities
are satisfied and that the VM predicts
$\chi_A = \chi_V = \frac{2}{3} \, T_c^2$ for $N_f = 2$,
which is in good agreement with the result obtained in the lattice 
simulation~\cite{Allton}.

In Ref.~\citen{HS:VD},
a prediction associated with the validity of 
vector dominance (VD) in hot matter was made:
As a consequence of including the intrinsic effect,
the VD is largely violated at the critical temperature.
This indicates that the assumption of the VD, which was made
in the analysis on the dilepton spectra carried out 
in hot matter such as in Ref~\citen{Rapp-Wambach:00},
may need to be weakened, at least in some amounts,
for consistently including the
effect of the dropping $\rho$ mass 
such as the one predicted by the 
Brown-Rho scaling~\cite{BR}
into the analysis.

In addition to the above predictions,
the pion velocity was studied
including the effect of Lorentz symmetry 
breaking~\cite{Sasaki:Vpi,HKRS:pvT},
which is reported in the write-up by Chihiro Sasaki~\cite{CS}.

\section*{Acknowledgements}
\vspace{-0.2cm}
I would like to thank the organizers of this workshop 
for giving me an opportunity to present my talk.
I am grateful to Dr. Youngman Kim, Professor Mannque Rho,
Dr. Chihiro Sasaki and Professor Koichi Yamawaki for collaboration
in the works on which this write-up is based.
This work is 
supported in part by the 21st Century COE
Program of Nagoya University provided by Japan Society for the
Promotion of Science (15COEG01).

\end{document}